\documentclass{article}

\usepackage{dcolumn}
\usepackage{epsfig}
\usepackage{cite}
\usepackage{amssymb,amsmath,bm,graphicx,multirow}
\usepackage{color}

\def\be{\begin{equation}}
\def\ee{\end{equation}}
\def\bea{\begin{eqnarray}}
\def\eea{\end{eqnarray}}
\def\bes{\begin{subequations}}
\def\ees{\end{subequations}}

\def\m{\mu}
\def\n{\nu}

\def\p{\partial}
\def\a{\alpha}
\def\b{\beta}
\def\t{\theta}

\def\11{1\hspace{-0.23cm}1}
\def\22{2\hspace{-0.23cm}2}
\def\33{3\hspace{-0.23cm}3}

\def\pp{p\hspace{-0.19cm}p}

\def\ph{\hat{p}}

\begin{document}
\begin{titlepage}
\begin{center}
\noindent{\large\textbf{One-electron atoms in Schwarzschild Universe: Bare and
electromagnetically dressed cases }}
\vspace{2\baselineskip}

Abolazl Jafari
\vspace{0.7cm}

\textit{
Department of Physics, Faculty of Science,\\
Shahrekord University, P. O. Box 115, Shahrekord, Iran\\
jafari-ab@sci.sku.ac.ir
}
\end{center}
\vspace{\baselineskip}
\begin{abstract}
The quantum mechanics of one-electron atoms in presence of external
electromagnetic fields is considered within Weber's framework.
The results by the earlier studies are extended in the sense
that for given source and field configurations the changes of the electromagnetic
potentials due to the curved background are included. The formulation is specialized to the case
with Schwarzschild background. The first corrections to the energy
levels for bare atom and Zeeman/Stark effects are calculated, exhibiting
possible changes in meaningful orders.
\end{abstract}

\vspace{1cm}

Pacs: 03.67.Mn, 73.23.-b, 74.45.+c, 74.78.Na


\end{titlepage}

\section{Introduction}
The behavior of quantum mechanical systems in the presence of gravitational fields
has been the subject of great number of research pieces. Among others, two leading
approaches are those by DeWitt \cite{dewitt} and Weber \cite{weber}. In the studies
based on DeWitt's approach, the general formulation of quantum mechanics for a
relativistic or non-relativistic system on a curved background is the main concern
\cite{parker1,parker2, pinto, mamwad}.
In the second based on Weber's, an interaction scheme between the quantum system and the
gravitational field is the guiding rule. In particular, in this approach the linearized
classical equations of motion of the test system/particle interacting with the gravitational
fields provide the basic ingredients to formulate the quantum theory \cite{weber,spelio, saha}.
Interestingly, these two approaches are not equivalent, and based on the different sequences
and orders of approximations being used in each approach, one may get different
results \cite{spelio, saha}.

Based on the DeWitt's approach, the formulation of Dirac particles on a curved
background is used to extract the first corrections in curvature to the
energy levels of one-electron atoms \cite{parker1,parker2, pinto, mamwad}.
In \cite{spelio, saha},
the interaction between gravitational waves and a charged test
particle is studied. While \cite{spelio, saha} falls within Weber's scheme,
it is shown that the sequence of linearizations used in the original version \cite{weber}
is not sufficient in case dealing with the charged particles in presence of external
fields.

The purpose of the present work is to extend the results for nonrelativistic charged
particles on a curved background.
In particular, within the Weber's framework,
we consider the case with one-electron atoms
in presence of additional external electromagnetic fields in the small curvature limit
to obtain the first corrections to the energy levels. Extending the results by
\cite{parker1,parker2, pinto}, for given source or field configurations,
the corrections due to curvature to the electromagnetic potentials as well as
and their effects on the energy levels are studied.
It will be seen that the obtained corrections to the
nuclei potential and the external fields due to curvature can
result in changes in meaningful orders of magnitude.
As a specific example, the corrections to the energy
levels of one-electron atom in the Schwarzschild metric is considered.

The scheme of the rest of this paper is the following. In section 2,
the basic notions of the formulation on curved background,
including the Riemann normal coordinate system is reviewed. In section 3,
the basic elements of the quantization procedure as well as the
the construction of Hamiltonian in the presence the electromagnetic potentials
based on Weber's approach  are presented.
In section 4 the formulation is specialized to the case for one-electron atom
in Schwarzschild background. In particular, for the case of bare atom and
the Zeeman/Stark effects the first corrections to the energy
levels are obtained. Section 5 is devoted to concluding remarks.

\section{Basic notions}
According to general relativity principles,
it is not possible to find a system of coordinates in curved space-time in which $\Gamma^{\gamma}_{\a\b}=0$ everywhere ($\a,\b,\gamma=0,1,2,3$).
However, one always can construct local inertial frames at a given event $P_0$,
in which free particles would move along straight lines locally.
As a consequence, it is possible to set  $\Gamma^{\gamma}_{\a\b}=0$
at least up to the first order of the Riemannian curvature.
As the constructed tangent space is similar to the Minkowski space-time,
a local inertial frame is defined for each given point $P_0$ of space-time
by the following equation for the metric
\begin{align}
g_{\a\b,\m}(P_0)=0.
\end{align}
The coordinates of such a frame are called the \textit{Riemann normal coordinate system} \cite{weber,misner,maggiore,inverno, weinberg}.
The metric components have the following forms in the Riemann coordinates up to the first order of Riemann's tensor ($i,j,\cdots=1,2,3$):
\begin{align} \label{metric}
g_{00}=-1-R_{0l0k}x^lx^k,
g_{0i}=g^{0i}=-\frac{2}{3}R_{0lik}x^lx^k,\cr
g_{ij}=\delta_{ij}-\frac{1}{3}R_{iljk}x^lx^k,
g^{00}=-1+R_{0l0k}x^lx^k,\cr
g^{ij}=\delta^{ij}+\frac{1}{3}R^{ij}_{lk} x^lx^k,
g=-1+\frac{1}{3}(R_{lk}-2R_{0l0k})x^lx^k.
\end{align}
Consequently, the affine connections (Christoffel multipliers) are found to be
\begin{align} \label{affine}&&
\Gamma^{0}_{00}=0,\ \ \ \Gamma^{0}_{ij}=\frac{1}{3}(R_{0ijk}+R_{0jik})x^k,\nonumber\\&&
\Gamma^{0}_{0i}=R_{0i0k}x^k,\ \ \ \Gamma^{i}_{jk}=\frac{1}{3}(R_{jikl}+R_{kijl})x^l,\nonumber\\&&
\Gamma^{i}_{0j}=R_{0kji}x^k,\ \ \ \Gamma^{i}_{00}=R_{0i0k}x^k.
\end{align}
The equation of the motion of a test particle in an arbitrary coordinate system reads
\begin{align} \label{equation of motion}
m\frac{d^2x^\mu}{d\tau^2}+m\,\Gamma^\mu_{\a\b}u^\a u^\b=F^\mu+\frac{q}{c}F_\mathrm{em}^{\mu\a}u^\b g_{\a\b},
\end{align}
with $u^\a=d x^\a/d\tau$ is the four vector velocity and $\tau$ is a proper time. $F^{\mu\alpha}_\mathrm{em}$ and $F^\mu$ stand for the space-time components of the
electromagnetic and other external forces acting on the particle, respectively.
The above equation of motion can be obtained in terms of the metric $ g_{\a\b}(x)$ and
the four vector electromagnetic potential  $A^\a$, usually adopted by the Lorentz gauge,
from the following Lagrangian \cite{misner,inverno, weinberg}
\begin{align} \label{initial lagrangian1}
\mathcal{L}=\frac{1}{2}m\dot{x}^\a g_{\a\b}(x)\dot{x}^\nu+\frac{q}{c}\dot{x}^\a g_{\a\b}(x)A^\b -\tilde{V}(x),
\end{align}
in which
\begin{align}
-\int^x F_\mu dx^\mu =\tilde{V}(x).
\end{align}

\section{Toward quantum system}
Here, using a set of assumptions and approximations, we develop the quantum mechanics
governing the dynamics of the test particle. As announced earlier, our approach is basically the
one by Weber's.

As we are considering nonrelativistic dynamics, it is assumed that the proper time
can be replaced by the coordinate time $x^0$, simply by $\tau\to c\,x^0$, by which
the equations of motion of the test particle for spatial directions become
\begin{align} \label{red eq motion}
m\ddot{x}_i+mR_{0i0j}x^j+\frac{1}{3}m(R_{jikl}+R_{kijl})x^l\dot{x}^j \dot{x}^k
=F^i +\Sigma_j\frac{q}{c}F^{\mathrm{em}}_{ij}\dot{x}_j +\frac{q}{c}F^{\mathrm{em}}_{i0},
\end{align}
Also and hereafter, we consider the cases for which we have:
\begin{align} \label{removed}
g_{0i}=0.
\end{align}
Many interesting cases, including the Schwarzschild metric, are of this type.
By these all we introduce the following Lagrangian
\begin{align} \label{initial lagrangian2}
\mathcal{L}=\frac{1}{2}m\dot{x}^i\delta_{ij}\dot{x}^j-\frac{1}{6}mR_{isjk}\dot{x}^ix^sx^k\dot{x}^j
+\frac{q}{c}\dot{x}^i\delta_{ij}A^j-\frac{q}{3c}R_{isjk}\dot{x}^ix^sx^kA^j
-\tilde{V}_\mathrm{eff}(x),
\end{align}
where
\begin{align}
\tilde{V}_\mathrm{eff}(x)=\tilde{V}(x)-\frac{1}{2}m c^2 g^{00} - \frac{q}{c} A_0.
\end{align}
The following is to show that the above Lagrangian produces
the desired equations of motion of the test particle according the Weber's picture.
Firstly, it is pointed out that the raising of the Lorentz indices
is done with the metric (\ref{metric}), namely,
$A_i=g_{ij}A^j\sim (\delta_{ij}-\frac{1}{3}R_{iljk}x^lx^k)A^j$.
Further, in the weak-field limit, the gravitational force is given by
$F^\a_\mathrm{grav}=-\frac{1}{2}m c^2 \frac{\p}{\p x_\a}g^{00}$ \cite{inverno, weinberg},
so we need to keep
the velocities independent term appearing in the Eq.(\ref{initial lagrangian1}).
Therefore, the gravitational parts will not be included in $\tilde{V}_\mathrm{eff}(x)$.
We mention that the metric is not explicitly a function of time.
For sake of simplicity, we set $m=1,\frac{q}{c}A_\m\rightarrow A_\m$:
\begin{align}
\frac{d}{d t}\frac{\p\mathcal{L}}{\p \dot{x}_k}-\frac{\p\mathcal{L}}{\p x_k}=
g^{ki}\ddot{x}_i
+\dot{g}^{ki}\dot{x}_i+\dot{g}^{ki}A_i+g^{ki}\dot{A}_i
-\frac{1}{2}\dot{x}_i\dot{x}_jg^{ij,k}
-g^{ij,k}\dot{x}_i A_j
\cr
-\dot{x}_ig^{ij} A_j^{,k}+\tilde{V}^{,k}
-\frac{1}{2}g^{00,k}-A_0^{,k}=0.
\end{align}
Using $\dot{g}^{ki}\dot{x}_i-\frac{1}{2}\dot{x}_i\dot{x}_jg^{ij,k}=\Gamma^{k}_{ij}\dot{x}^j\dot{x}^i$,
and $\Gamma^k_{00}=-\frac{1}{2}g_{00}^{,k}$, and
referring to (\ref{removed}), $\Gamma^k_{0j}=0$, $A_0^{,k}=F^{0k}g_{00}$, and
\begin{align}
\dot{g}^{ki}A_i+g^{ki}\dot{A}_i-g^{ij,k}\dot{x}_i A_j-\dot{x}_ig^{ij} A_j^{,k}
=g^{kj,i}\dot{x}_iA_j+g^{kj}\dot{x}_iA_j^{,i}
\cr
-g^{ij,k}\dot{x}_i A_j-\dot{x}_ig^{ij} A_j^{,k}
=\p^i(g^{kj}A_j)\dot{x}_i-\p^k(g^{ij}A_j)\dot{x}_i
\cr
=A^{k,i}\dot{x}_i-A^{i,k}\dot{x}_i=F^{ki}\dot{x}^jg_{ji}.
\end{align}
It is seen that the Lagrangian (\ref{initial lagrangian2}) can produce the equation of the motion
(\ref{red eq motion}). The Hamiltonian of the system can be constructed easily using the Legendre transformation. The conjugate momentum is as follows:
\begin{align}
p_k=m\dot{x}^i\delta_{ik}-\frac{1}{3}mR_{kjil}x^jx^l\dot{x}^i+\frac{q}{c} A^j\delta_{jk}-\frac{q}{3c}R_{kjil}x^jx^lA^i,
\end{align}
by which,
\begin{align} &&
m\dot{x}^i(\delta_{ik}-\frac{1}{3}R_{kjil}x^jx^l)=p_k-\frac{q}{c}\delta_{ki} A^i+\frac{q}{3c}R_{kjil}x^jx^lA^i.
\end{align}
By the set of coordinates and their conjugate momenta, the Hamiltonian comes to the form
\bea \label{hamiltoni00}
\mathcal{H}&=&\dot{x}^ip_i-\mathcal{L}
\nonumber\\&=&
\frac{1}{2}m\dot{x}^i\delta_{ij}\dot{x}^j-\frac{1}{6}mR_{ikjl}\dot{x}^ix^kx^l\dot{x}^j+\tilde{V}_\mathrm{eff}
\nonumber\\&=&
\frac{1}{2}\dot{x}^i(m\dot{x}^j(\delta_{ij}-\frac{1}{3}R_{ikjl}x^kx^l))+\tilde{V}_\mathrm{eff}
\nonumber\\&=&
\frac{1}{2}(\dot{x}_s(\delta^{si}+\frac{1}{3}R^{si}_{qr}x^qx^r))(m\dot{x}^j(\delta_{ij}-\frac{1}{3}R_{ikjl}x^kx^l))+\tilde{V}_\mathrm{eff}
\nonumber\\&=&
\frac{1}{2m}(p^k-\frac{q}{c}\delta^{ik} A_i-\frac{q}{3c}R^{ki}_{jn}x^jx^nA_i)(p_k-\frac{q}{c}\delta_{ks} A^s+\frac{q}{3c}R_{klsr}x^lx^rA^s)
\nonumber\\&&+
\tilde{V}_\mathrm{eff}
\nonumber\\&=&
\frac{1}{2m}p_kp^k-\frac{q}{2mc}p^k\delta_{ki}A^i-\frac{q}{2mc}A_i\delta^{ik}p_k\nonumber\\&&
+\frac{q}{6mc}R_{kjil}p^kx^jx^lA^i-\frac{q}{6mc}R^{ki}_{jl}x^jx^lA_ip_k+\frac{q^2}{2mc^2}A_i\delta^{ik}
\delta_{ki}A^i\nonumber\\&&
-\frac{q^2}{6mc^2}R_{kjil}\delta^{ik}A_ix^jx^lA^i+\frac{q^2}{6mc^2}R^{ki}_{jl}x^jx^lA_i\delta_{ki}A^i
+\tilde{V}_\mathrm{eff},
\eea
or equivalently, it can be reduced to the following
\begin{align} \label{hamiltoni01}
\mathcal{H}=
\frac{1}{2m}\mathbf{p}^2-\frac{q}{2mc}(p^k\delta_{ki}A^i+A_i\delta^{ik}p_k)
\cr+\frac{q}{6mc}
(R_{kjil}p^kx^jx^lA^i-R^{ki}_{jl}x^jx^lA_ip_k)
+\tilde{V}_\mathrm{eff}
+O(\mathbf{A})^2,
\end{align}
where, the Coulomb potential appearing in $\tilde{V}_\mathrm{eff}(x)$, is given by the Maxwell equations in the curved background.

Here we assume that the solutions to the Maxwell equations for
the potentials $A_i$'s are subjected to the Coulomb gauge, by which
\begin{align}
\ph^k\delta_{ki}\hat{A}^i=\hat{A}_i\delta^{ik}\ph_k,
\end{align}
where the symbol $\hat{~}$ indicates the operator forms of the
variables are being used.
It is easy to check that due to the diagonal form of
Riemann's curvature tensors of the Schwarzschild universe, the above form is possible.

In passing to quantum theory, the classical values are replaced
by their operator counterparts.
Due to terms involving coordinates and momenta, one encounters the known problem
of ordering ambiguity. Here we exploit the rising of the Latin indices to construct
the symmetrical Weyl ordering, by which
the Hamiltonian (\ref{hamiltoni00}) comes to the form
\begin{align} \label{h0}
\hat{\mathcal{H}}&=\frac{1}{2m}\hat{\mathbf{p}}^2-\frac{q}{mc}\hat{\mathbf{p}}\cdot\hat{\mathbf{A}}+
\frac{q^2}{2mc^2}\hat{\mathbf{A}}\cdot\hat{\mathbf{A}}
+\frac{q}{3mc} R_{ijij}\hat{p}^i\hat{x}^j\hat{x}^j\hat{A}^i
\cr
&+\frac{q}{9mc} R_{ijji}(3\hat{p}^i\hat{x}^j\hat{x}^i\hat{A}^j+2\imath\hbar A^j\delta^{ii}+\imath\hbar x^i\p^i A^j)
+\hat{\tilde{V}}_\mathrm{eff}(\hat{x})+O(\mathbf{A})^2.
\end{align}

In order to consider all of the corrections in first order of curvature, the electromagnetic
potentials should also be re-calculated. The Maxwell equations in the curved background take
the following form
\begin{align}
g^{\a\b}\nabla_\a\nabla_\b A_\m-R_{\m}^{\n}A_\n=-4\pi J_\m,
\end{align}
for which, by applying the local coordinates introduced earlier, we have \cite{parker1, parker2}:
\begin{align} \label{parker}
\delta^{ij}\p_i\p_j A_0+\frac{1}{3}R_{iljk}x^kx^l\p^i\p^j A_0+\frac{5}{3}R_{i00j}x^j\p^iA_0
+2R^k_{i0j}x^j\p^i A_k
\cr
-\frac{2}{3}R^i_jx^j\p_i A_0=-4\pi J_0,\\
\delta^{ij}\p_i\p_j A_m+\frac{1}{3}R_{iljk}x^kx^l\p^i\p^j A_m-\frac{2}{3}R^\m_m A_\m
-\frac{1}{3}R^l_{00m}A_l
\cr
+\frac{2}{3}\delta^{ij}(R^\a_{mik}+R^\a_{imk})x^k\p_j A_\a
-\frac{1}{3}R^i_{00j}x^j\p_i A_m
\cr
-\frac{2}{3}R^i_{j}x^j\p_i A_m=-4\pi J_m.
\end{align}
Using the perturbative expansion $A_\a=A^0_{\a}+A^1_\a+O(R)^2$,
for the source of nucleus $J_0=-Q/r$ and $J_i=0$, one finds the following for
the electromagnetic potentials:
\bea
&A^\mathrm{nucl.}_0=-Qr^{-1}+\frac{1}{12}Q(R+4R_{00})r+\frac{1}{12}Q(3R^0_{j0k}-R_{jk})x^jx^kr^{-1},
\\
&A^\mathrm{nucl.}_m=\frac{1}{2}Q R_{0m}r+\frac{1}{6}Q R^0_{jmk}x^jx^kr^{-1}.
\eea
Assuming that the electromagnetic potentials have two parts, corresponding the one by
nucleus of one-electron atom and the one by the external sources (as in Zeeman and Stark effects),
we use the following replacement:
\begin{align}
A_\a\to A^\mathrm{nucl.}_\a+A_\a,
\end{align}
in which the second term is responsible for the potential by the external sources.
The latest Hamiltonian takes the form:
\begin{align}
\mathcal{\hat{H}}\simeq\frac{1}{2m} \hat{\pp}^2
- Qe\hat{r}^{-1}
 +\frac{1}{12}Qe((R+4R_{00})\hat{r}+(3R^0_{j0k}-R_{jk})\hat{x}^j\hat{x}^k)\hat{r}^{-1}
\cr
+\frac{1}{2}m R_{0s0k}\hat{x}^s\hat{x}^k
+ \frac{e^2}{2mc^2}\hat{\mathbf{A}}\cdot\hat{\mathbf{A}}-\frac{e}{mc}\hat{\mathbf{p}}\cdot\hat{\mathbf{A}}
+\frac{e}{3mc}
R_{ijij}\hat{p}^i\hat{x}^j\hat{x}^j\hat{A}^i
\cr
+\frac{e}{9mc}
R_{ijji}(3\hat{p}^i\hat{x}^j\hat{x}^i\hat{A}^j+2\imath\hbar A^j\delta^{ii}+\imath\hbar x^i\p^i A^j).
\end{align}
The above Hamiltonian consists of local coordinates and is valid for the non-covariant observer as well. As a consequence, the results
from the theory only can be interpreted in a local framework based on Riemann normal coordinates.
As is evident, these two are identical in Schwarzschild background without external
electromagnetic fields.

\section{Quantum theory in Schwarzschild background}
As an application of the quantum theory developed in previous
section, here we consider the background by the static Schwarzschild solution
\begin{align}
ds^2=-c^2\left(1-\frac{2GM}{c^2r}\right)dt^2+\left(1-\frac{2GM}{c^2r}\right)^{-1}dr^2
\cr
+r^2(d\t^2+\sin^2{\t}\,d\phi^2),~~~~~~~~~~~~~~~~~~~~~~
\end{align}
in which $r$, $\theta$ and $\phi$ are representing the spherical coordinates.
The nonvanishing $R_{\m\n\a\b}$ are the relevant spatial components of curvature tensor
in the spherical coordinates given by
\begin{align} \label{schwarzschild components}
&&
R^1_{010}=-2R^2_{020}=-2R^3_{030}=\frac{2GM(2GM-rc^2)}{c^4r^4},\nonumber\\&&
R^3_{232}=\frac{2GM}{c^2r},~~~~
R^3_{131}=R^2_{121}=\frac{GM}{r^2(2GM-rc^2)}.
\end{align}
Evidently, the components of curvature tensor would get arbitrary smaller values
in the limit $r\to\infty$. This fact is the basis to use the results in the previous section which
are valid in the small curvature limit. Therefore, for one-electron atoms
sufficiently far from the origin the quantum theory developed in Sec.~3 is applicable.

\subsection{Bare one-electron atom}
Setting $A_\a^0=A_a^1=0$, we have
\begin{align} \label{ffree atom}
\mathcal{H}=\mathcal{\hat{H}}_0+
\frac{1}{2}m R_{0i0j}\hat{x}^i\hat{x}^j+
\frac{1}{4}QeR^0_{i0j}\hat{x}^i\hat{x}^jr^{-1},
\end{align}
where $\mathcal{\hat{H}}_0$ stands for the unperturbed Hamiltonian
of the one-electron atom. The second term in above represents
the direct effect of gravitational field on the energy, firstly considered
by \cite{parker1,parker2} and used by \cite{pinto} to obtain the
corrections to energy levels and the transition rates. The third term obtained in
above is evidently originated from the correction to the nuclei potential due
to the curvature. As we will see this added term would give comparable changes
in the energy levels. The typical radius of curvature
should be as small as $D\sim 10^{-3}~\mathrm{cm}$, by which
at the nonrelativistic limit the perturbation would give
larger corrections than the relativistic fine structure \cite{parker1,parker2}.
The following relations are useful in the subsequent calculations:
\begin{align*}
\frac{x^2}{r^2}&=\frac{\sqrt{4\pi}}{3}Y_{0,0}(\theta,\phi)-\frac{1}{6}\sqrt{\frac{16\pi}{5}}Y_{2,0}(\theta,\phi)
+\frac{1}{4}\sqrt{\frac{32\pi}{15}}(Y_{2,2}(\theta,\phi)+Y_{2,-2}(\theta,\phi))
\cr
\frac{y^2}{r^2}&=\frac{\sqrt{4\pi}}{3}Y_{0,0}(\theta,\phi)-\frac{1}{6}\sqrt{\frac{16\pi}{5}}Y_{2,0}(\theta,\phi)
-\frac{1}{4}\sqrt{\frac{32\pi}{15}}(Y_{2,2}(\theta,\phi)
+Y_{2,-2}(\theta,\phi)),
\cr
\frac{z^2}{r^2}&=\frac{\sqrt{4\pi}}{3}Y_{0,0}(\theta,\phi)+\frac{1}{3}\sqrt{\frac{16\pi}{5}}Y_{2,0}(\theta,\phi),
\end{align*}
also
\begin{align*}
\int_0^{\pi}d\theta\sin{\theta}\int_0^{2\pi}d\phi\ Y^\star_{lm}(\theta,\phi)Y_{l_1m_1}(\theta,\phi)Y_{l_2m_2}(\theta,\phi)
\cr=
\sqrt{\frac{(2l_1+1)(2l_2+1)}{4\pi(2l+1)}}C^{l_1l_2,l}_{00,0}C^{l_1l_2,l}_{m_1m_2,m},
\end{align*}
where $C^{l_1l_2,l}_{m_1m_2,m}$ stands for Clebsch-Gordan coefficients.
In what follows we represent the one-electron atom states as usual
\begin{align}
|\psi\rangle=| n\,l\,m_l\rangle,
\end{align}
where $n,l$ and $m_l$ are the relevant quantum numbers.
By these the first-order correction to energy of $S$-states  ($l=0$) is given by
\begin{align}
\langle H^1_{S}\rangle=\frac{1}{4}QeR^0_{i0j}\int d^3 x\ \psi^\star_{n00}(x)x^ix^jr^{-1}\psi_{n00}(x).
\end{align}
Using the given form of $R_{0i0j}$ by Eq.(\ref{schwarzschild components}), we readily have
\begin{align}
R_{0101}x^2+R_{0202}y^2+R_{0303}z^2=R_{0303}(r^2-3x^2),
\end{align}
by which for the $S$-states
the first correction to the energy by the correction to the nuclei
potential by curvature vanishes.
By this the present model coincides with those by Parker and Pinto's
for $S$-states \cite{parker1,parker2,pinto}.
However, the situation is different for the $P$-states ($l=1$).
To calculate the $P$-states ($l=1$) the diagonalization of
the degenerate block of the Hamiltonian is required.
Due to the electric quadrupole transition selection rules,
we have $\Delta l=0,\pm 2$ and $\Delta m=0,\pm 1,\pm 2 $.
With similar calculation for $P$-states,
One finds the matrix elements of the shifted Hamiltonian
\begin{align}
H^1_{m m' }&=\frac{1}{4}QeR_{0303}
\left(\langle\hat{r}\rangle_{_{n,1}}-3\langle \frac{\hat{x}^2}{\hat{r}}\rangle\right)
\cr
&=\frac{1}{4}QeR_{0303}\langle r\rangle_{_{n,1}}\left(
C^{21,1}_{00,0}C^{21,1}_{0 m' ,m}
\right.
\cr &~~~~
-\sqrt{\frac{3}{2}}C^{21,1}_{00,0}(C^{21,1}_{2 m' ,m}
+C^{21,1}_{-2 m' ,m})\Big),
\end{align}
where $m$ and $ m' $ take value $0, \pm 1$, corresponding to the $P_x$, $P_y$ and $P_z$ orbits, respectively.  By setting $\beta=\frac{1}{10}QeR_{0303}\frac{\hbar^2}{me^2}(3n^2-2)$, and the diagonal form of
$R_{0i0j}$, the explicit form of the matrix is found to be
\begin{align}
H^{1}_{m m' }=\left(
                     \begin{array}{ccc}
                       -\frac{\b}{4} & 0 & \frac{3\b}{4} \\
                       0 & \frac{\b}{2} & 0 \\
                       \frac{3\b}{4} & 0 & -\frac{\b}{4} \\
                     \end{array}
                   \right).
\end{align}
By the corresponding  eigenvalue equation:
\begin{align}
\mathrm{Det}\Big{(}H^1_{m m' }-E^1\delta_{m m' }\Big{)}=0,
\end{align}
the following values are obtained for the corrections,
\begin{align}
\label{ep}
E^1_{P}=\frac{\b}{2},-\b.
\end{align}
As mentioned earlier, the correction to the nuclei potential is absent in \cite{pinto}.
As a consequence, in \cite{pinto} the $S$-states, in agreement with the present model
would not get corrections at first order.
It would be useful to compare the result for the $P$-states.
By the last expression
in above, we would get
\begin{align}
\frac{E^1_{new}}{E^1_{Pinto}}\sim
10^{-1}\frac{Qe\langle r\rangle}{m\langle r^2\rangle}\propto 10^{-1}\frac{e^2(3n^2-l(l+1))}{2ma_0n^3(2l+1)},
\end{align}
in which,
$E^1_{new}$ and $E^1_{Pinto}$ are the first corrections due to the corrections
to the nuclei potential by the present model, and the curvature by \cite{parker1, pinto}.
As is evident, the corrections by the additional term in (\ref{ep})
is not negligible and have to be considered.

\subsection{The Normal Zeeman Effect}
Here we consider the effect of an external uniform magnetic field on the
energy levels of the one-electron atom, which is, in absence of
spin effects, known as normal Zeeman effect.
For the nonzero uniform magnetic background field as $\mathbf{B}=B_0\,\hat{k}$,
due to the specific form of the Riemann tensor, it is easy to see that
there is no change in the correction to the scalar potential $A_0$ of previous
section, and we have:
\begin{align}
A^1_0=\frac{1}{4}QR^0_{j0k}x^jx^kr^{-1}.
\end{align}
In the Coulomb gauge, the unperturbed components
of the vector potential are given as
\begin{align}
A^0_i=-\frac{B_0}{2}\epsilon_{ij3}x^j.
\end{align}
The components of the correction to the potential, $A^1_i$,  satisfy
\begin{align}
\p^2 A^1_i=\frac{B_0}{6}(3R^2_{002}+2R^1_{212})\epsilon_{ij3}x^j,
\end{align}
by which, using $R_{0101}+R_{0202}+R_{0303}=0$, we have
\begin{align}
\mathbf{A}^1=\frac{B_0}{6}(3R^2_{002}+2R^1_{212})
(\frac{1}{3}y^3-yx^2,~-\frac{1}{3}x^3+xy^2,~0),
\end{align}
by which at the first order in curvature,
\begin{align}
\mathbf{A}=\left(
  \begin{array}{c}
    -\frac{B_0}{2}y+\frac{B_0}{6}(3R^2_{002}+2R^1_{212})(\frac{1}{3}y^3-yx^2)\\
     \frac{B_0}{2}x-\frac{B_0}{6}(3R^2_{002}+2R^1_{212})(\frac{1}{3}x^3-xy^2) \\
    0 \\
  \end{array}
\right).
\end{align}
By these all the operator form of the Dewitt's Hamiltonian of the one-electron atom for the
normal Zeeman effect takes the form
\begin{align} \label{Dewitt Hydrogen atom zeeman}
\mathcal{\hat{H}}^{\emph{Zeeman}}_{Perturbed}&\simeq
-\frac{e B_0}{2mc}\mathcal{\hat{L}}_{\emph{z}}
-\frac{eB_0}{6mc}(3R^2_{002}+2R^1_{212})(\ph_x (\frac{1}{3}\hat{y}^3-\hat{y}\hat{x}^2)
\nonumber\\&
-\ph_y(\frac{1}{3} \hat{x}^3-\hat{x}\hat{y}^2))+O(B_0)^2,
\end{align}
in which the first two terms represent the unperturbed Hamiltonian. The
above expression is valid for the so-called weak field, or $B_0\lesssim 10^{-2}$ G.
In fact the last term in above is the result of corrections to
the potential,
can be calculated the contribution of the correction related to the Zeeman effect, it is as follows:
\begin{align}
\Delta E^{Zeeman}_{Perturbed}&=&
\Delta E^{Zeeman}_0+\Big{(}\frac{e B_0}{18mc}(3R^2_{002}+2R^1_{212})(n^2\frac{5n^2+1-3l(l+1)}{2}a_0^2)
\nonumber\\&&\times m_l\hbar(1-C^{2l,l}_{00,0}C^{2l,l}_{0m_l,m_l})\Big{)}.
\end{align}
Evidently, there is nonzero corrections due to the second part
for the $P$-states.
Whereas, the Weber's Hamiltonian of the one-electron atom for the
normal Zeeman effect takes the form
\begin{align} \label{Weber Hydrogen atom zeeman}
\mathcal{\hat{H}}^{\emph{Zeeman}}_{Perturbed}&\simeq
-\frac{e B_0}{2mc}\mathcal{\hat{L}}_{\emph{z}}
+\frac{eB_0}{6mc}\epsilon^{ji}R_{ijij}\hat{p}^i\hat{x}^j\hat{x}^j\hat{x}^j
\nonumber\\&
+\frac{eB_0}{18mc}R_{ijji}\epsilon^{ij}(3\hat{p}^i\hat{x}^j\hat{x}^i\hat{x}^j+2\imath\hbar \hat{x}^i\delta^{ii}+\imath\hbar\hat{x}^i\delta^{jj})
+O(B_0)^2,
\end{align}
in which, 
it can be transformed to a more useful form by the
identity
\begin{align}
\ph_x \hat{y}^3-\ph_y \hat{x}^3=\mathcal{\hat{L}}_{\emph{z}}(\hat{r}^2-\hat{z}^2)+\ph_y\hat{x}\hat{y}^2-\ph_x\hat{y}\hat{x}^2.
\end{align}
By the above, the contribution of the correction related to the Zeeman effect
can be calculated. Following \cite{pinto} and in a semi-classical point of view,  we can
calculate the energy levels based on the Bohr's quantization procedure.
In fact, the stability of motion of the one-electron atom
in the $xy$ plane does not last long.
However, after neglecting the effect of the last correction,
the Bohr radius is definable.
One can choose the
orientation of the spatial axes of the normal coordinates
such that $z\sim y'$, $y\sim x'$ and $x\sim z'$.
Now, if we restrict ourselves to circular orbits in the ${x'}{y'}$
plane and by assuming the presence of a magnetic field
in $z$ (radial) direction and with ${\rho'}=\sqrt{{x'}^2+{y'}^2}$, the equation of motion can be
shown to be
\begin{align}
m\frac{{v'}^2}{{\rho'}}=\frac{e{v'}B_0}{c}+\frac{Qe}{{\rho'}^2}+mR_{0202}{\rho'}+\frac{Qe}{4}R_{0202},
\end{align}
by the Bohr quantization condition ($m\rho v_n=n\hbar$) and $R=R_{0202}$, we get
\begin{align}
n^2-\frac{eB_0}{c\hbar}n{\rho'}^2=\frac{zme^2}{\hbar^2}{\rho'}+\frac{mQeR}{4\hbar^2}{\rho'}^3+\frac{m^2R}{\hbar^2}{\rho'}^4,
\end{align}
or
\begin{align}
(n-\frac{eB_0}{2c\hbar}{\rho'}^2)^2=\frac{mQe}{\hbar^2}{\rho'}+\frac{mQeR}{4\hbar^2}{\rho'}^3
+(\frac{m^2R}{\hbar^2}
+\frac{e^2B^2_0}{4c^2\hbar^2}){\rho'}^4.
\end{align}
Comparing above with the similar one by \cite{pinto}, the term proportional
to $\rho'^3$ is new. Therefore, although there is no change in the Landau-Bohr
radius obtained in \cite{pinto}, however a new radius can be defined
by this new term, namely
\begin{align}
r_a=\left(\frac{4\hbar^2}{3mQeR}\right)^{{1}/{3}}.
\end{align}
The obtained radius goes to the infinity when $R=0$, leading to the
motion on the straight line.

\subsection{The Stark Effect}
This energy shift of atomic levels
in presence of an external uniform electric field is known as the Stark effect.
On this topic, the Hamiltonian describing the Stark effect is encountered with a technical agreement between the DeWitt and Weber approaches.
The deformed Maxwell equations are used to reproduce of the potential and 
redefine the distribution of electrical charges in order to product an uniform electric fields in the presence of the gravitational background.
To be specific and for the case of Weber's method, we assume that 
$J_0=Q\delta(\vec{r}_-)-Q\delta(\vec{r}_+)$, with $\vec{r}_\pm=\vec{r}\pm\vec{R}$.
So that, if the size of $\vec{R}$ is infinite then the electric field will be uniform.
However, according to the Eqs.(\ref{parker}), the scalar potential comes to the form:
\begin{align} 
A_0=A^0_0-\frac{1}{4}E_0R^0_{i0j}x^ix^jz,
\end{align}
where, 
\begin{align} 
A^0_0=\frac{Q}{r_-}-\frac{Q}{r_+}
=
Q\Sigma_0^\infty\frac{r^l}{R^{l+1}}(P_{l}(\cos{(\gamma)}-P_{l}(-\cos{(\gamma)})=\frac{2Q}{R^2}r\cos{(\gamma)}\mid_{\frac{2Q}{R^2}=E_0},
\end{align}
where, $\gamma$ is a angle between $\vec{r}$ and $\vec{R}$.
So, by setting $\vec{R}=R\hat{k}$, the potential by which the uniform electric filed in curved background is produced
is given by
\begin{align}
A_0=-E_0r\cos(\theta)-\frac{E_0}{4}R^0_{i0j}x^ix^jr\cos(\theta)=-E_0z-\frac{E_0}{4}R^0_{i0j}x^ix^jz.
\end{align}
Hence, in Schwarzschild background, the perturbed Hamiltonian for
the Stark effect is given by
\begin{align} \label{Hydrogen atom stark}
\mathcal{\hat{H}}^{\emph{Stark}}_{perturbed}&\simeq&\mathcal{H}_0
+\frac{1}{4}QeR^0_{i0j}
\hat{x}^i\hat{x}^j\hat{r}^{-1}
+\frac{1}{2}m R_{0i0j}\hat{x}^i\hat{x}^j+
eE_0\hat{z}+e\frac{E_0}{4}R^0_{i0j}\hat{x}^i\hat{x}^j\hat{z}.
\end{align}
Due to the parity, the correction to energy of $S$-states
by the term $e\frac{E_0}{4}R^0_{i0j}x^ix^jz$ vanishes.

\section{Concluding remarks}
The results for nonrelativistic charged particles on a curved background
are extended. In particular, within the Weber's framework,
we consider the case with one-electron atoms
in presence of additional external electromagnetic fields in the small curvature limit
to obtain the first corrections to the energy levels. Extending the results by
\cite{parker1,parker2, pinto}, for given source or field configurations,
the corrections due to curvature to the electromagnetic potentials as well as
and their effects on the energy levels are studied.
It will be seen that the obtained corrections to the
nuclei potential and the external fields due to curvature can
result in changes in meaningful orders of magnitude.
As a specific example, the corrections to the energy
levels of one-electron atom in the Schwarzschild metric is considered.

\section{Acknowledgments}
The author thanks professor Amir H. Fatollahi for guidance and advice in the conduct of 
this research.
Also, the author thanks the Shahrekord University for support of this research grant fund.
\newline

\end{document}